\documentclass{aa}  

\usepackage{graphicx}
\usepackage{multirow}
\usepackage{tikz}
\usepackage[separate-uncertainty=true, retain-unity-mantissa = false]{siunitx}

\DeclareSIUnit\angstrom{\text {Å}}
\DeclareUnicodeCharacter{0301}{\'{e}}

\newcommand{\figref}[1]{\figurename~\ref{#1}}
\newcommand{\secref}[1]{Sect.~\ref{#1}}

\newcommand{\epi}{\textsc{epimetheus}}
\newcommand{\pa}{\widehat{\mathrm{PA}}}

\usepackage[varg]{txfonts}

\usepackage{xcolor}
\definecolor{blue_dark}{rgb}{0.2, 0.2, 0.6}
\definecolor{blue_green}{rgb}{0.1, 0.4, 0.3}
\usepackage[hidelinks,colorlinks=true,linkcolor={blue_dark},citecolor={blue_dark}, urlcolor={blue_green},filecolor={blue_dark}]{hyperref}
\usepackage{orcidlink}

\begin{document} 

   \title{EPIMETHEUS: Effective PSF inferred from mosaics}

   \author{Pietro Benotto\orcidlink{0009-0004-7602-0160}\inst{1,2}\fnmsep\corrauth{pietro.benotto@inaf.it}
   \and
   Benedetta Vulcani\orcidlink{0000-0003-0980-1499}\inst{1}
   \and
   Vittoria Altomonte\orcidlink{0009-0000-7860-7556}\inst{2}}

   \institute{
        \inst{1} INAF, Osservatorio Astronomico di Padova, Vicolo dell’Osservatorio 5, 35122, Padova, Italy\\
        \inst{2} Dipartimento di Fisica e Astronomia `Galileo Galilei', Università di Padova, Vicolo dell'Osservatorio 3, 35122 Padova, Italy\\
    }

   \date{Received 13 April 2026 / Accepted dd September 2026}
   
   \abstract{Accurate knowledge of the point spread function (PSF) is essential for a wide range of astronomical applications, from astrometry to precision photometry. However, deriving a robust average effective PSF (ePSF) from astronomical mosaics is particularly challenging due to the superposition of exposures taken at different epochs, under different observing conditions, and with varying position angles (PAs). To address this issue, we present \epi, an open-source Python library designed to manage the entire ePSF construction workflow, from initial star selection to final model generation. Built with a modular architecture, \epi\ avoids black-box methodologies, ensuring full transparency and giving the user complete control over the stellar sample to use for the ePSF building to meet specific scientific requirements. The software can process multiple images or mosaics simultaneously to retrieve the ePSF, a crucial feature when individual fields lack a sufficient number of bright, unsaturated stars. Candidate stars can be automatically filtered based on orientation and light concentration, with options for manually excluding stars through a semi-automatic visual inspection. Although \epi\ is specifically optimised to handle the roto-translation complexities inherent to multi-epoch mosaics, it is versatile and can be equally applied to derive ePSFs from simpler, single-orientation images where rotational corrections are not required. Finally, to demonstrate the software's capabilities and provide an immediate resource, we release empirical ePSFs for standard HST and JWST filters across the COSMOS, UDS, Abell 2744, and GOODS-S fields. }
   \authorrunning{P. Benotto et al.}
   \maketitle

\section{Introduction}\nolinenumbers
In observational astronomy, knowing the instrumental point spread function (PSF) is essential to perform accurate measurements of both the astrometrical and photometrical information of the objects analysed.

For pixelated data, especially in the undersampled regime, the relevant quantity is the effective PSF (ePSF), which includes the detector pixel response. Dithered observations are particularly useful in this context, because they allow one to reconstruct a better-sampled effective PSF and to preserve both photometric and astrometric fidelity when combining undersampled images \citep{Lauer:1999, Anderson:2000, Fruchte:1998}.

A common procedure, in particular when the region of interest exceeds the instrument field of view, is to make use of mosaics (i.e. large-scale images obtained by combining multiple, partially overlapping exposures), often taken at different epochs and telescope position angles (PA). The result is a higher signal-to-noise ratio (S/N) image with a wider field of view. 

Accurately modelling the PSF across one mosaic field is a complex task, as it depends on both the position of the source on the detectors in each exposure composing the mosaic \citep[e.g.,][]{Stetson:1994, Tomaney:1996, Anderson:2000, Fusco:2000, Jee:2007, Nardiello:2022}, and the overlap of the exposures taken at different PAs. The problem is further complicated by the fact that the PSF can vary from exposure to exposure because of changes in focus, optical configuration, detector properties, and observing conditions \citep{Rhodes:2007, Jee:2007}. For ground-based observations, the PSF can also vary because of atmospheric conditions, and these temporal variations can introduce substantial residual systematics \citep{Hoekstra:2003, Marois:2006, Lazorenko:2009}.

Given these intrinsic difficulties, and since a point-by-point ePSF modelling is generally beyond the scope of most scientific applications based on mosaics, it is common practice to derive an average ePSF across the field or across subsets of exposures with similar properties \citep{Makovoz:2005, Rhodes:2007, Jee:2007, Whitaker:2019, Weaver:2024, Merlin:2024, Sarrouh:2025}. This approximation is often sufficient for source extraction and many photometric applications, although its adequacy ultimately depends on the scientific requirements and on the amplitude of the spatial and temporal PSF variations within the mosaic.

There is no unique methodology for deriving such an average ePSF, and the approaches adopted in the literature vary significantly: some studies manually select and rotate individual stars to a common orientation to derive a unique ePSF across multiple mosaics \citep{Merlin:2024}; others stack all suitable stars together regardless of their orientation \citep{Weaver:2024}; yet others combine an extracted ePSF for the core with a modelled PSF to better constrain the wings \citep{Sarrouh:2025}.

The heterogeneity of these methodologies is partly due to the lack of a unified software that guides the user through the complete ePSF extraction workflow --- from initial star selection to final model construction --- while offering the flexibility required to address the specific challenges posed by mosaic data. For example, \textsc{PSFEx}\footnote{Throughout this manuscript, software and package names are displayed in {\string\sc} while functions, classes, and code snippets in {\string\tt}.} \citep{Bertin:2013} is not implemented in Python, and it is hardly customizable, while \textsc{photutils} \citep{photutils} provides the \texttt{EPSFBuilder} function, which lacks full user control over the selection of stars used to construct the ePSF and does not account for PSF rotation arising from multiple PAs.
Furthermore, existing software lacks the capability to jointly analyse multiple images together, combining their stellar samples to overcome the limited number of stars available in any individual field while still producing a dedicated ePSF for each image.

In this work, we present and publicly release \epi\footnote{\url{https://github.com/pietrobenotto/epimetheus}} (Effective Psf Inferred from Mosaics using an Estimation Technique for Heterogeneous Exposures to get a Unique PSF across the Sensor field), a Python-based software designed to extract ePSFs from input data ranging from individual exposures to complex mosaics. \epi\ was originally conceived to operate on stacked extragalactic mosaics, where only a handful of faint stars are typically available, and it automatically measures and accounts for the rotation of each detected star in the input images, allowing the user to either stack the entire star sample into a composite model or select only those sharing a common orientation. The software can also operate on multiple science images simultaneously, using the combined stellar sample to maximise the ePSF S/N while producing field-specific models aligned with the dominant orientation of each image. To minimise information loss, all geometric transformations (rotation, translation, and resampling) are combined into a single final interpolation step, mapping the initial cutout directly onto the final stacked ePSF frame. Despite being designed with this challenging
scenario in mind, \epi\ is equally well suited to single-sensor frames, even for star-crowded fields, where the rotation procedure is not strictly necessary: in this case, the software provides a fast and customizable procedure to extract ePSFs, making it a versatile tool across a wide range of observing setups.

The paper is organised as follows: \secref{sec:epi} describes the features and the structure of the \epi\ library, \secref{sec:discussion} discusses the advantages of \epi\ and its limitations, while \secref{sec:conclusions} reports the key points of \epi\ and of this work.

\section{Software description}\label{sec:epi}

\begin{figure*}[htbp]
    \includegraphics[width=1\textwidth]{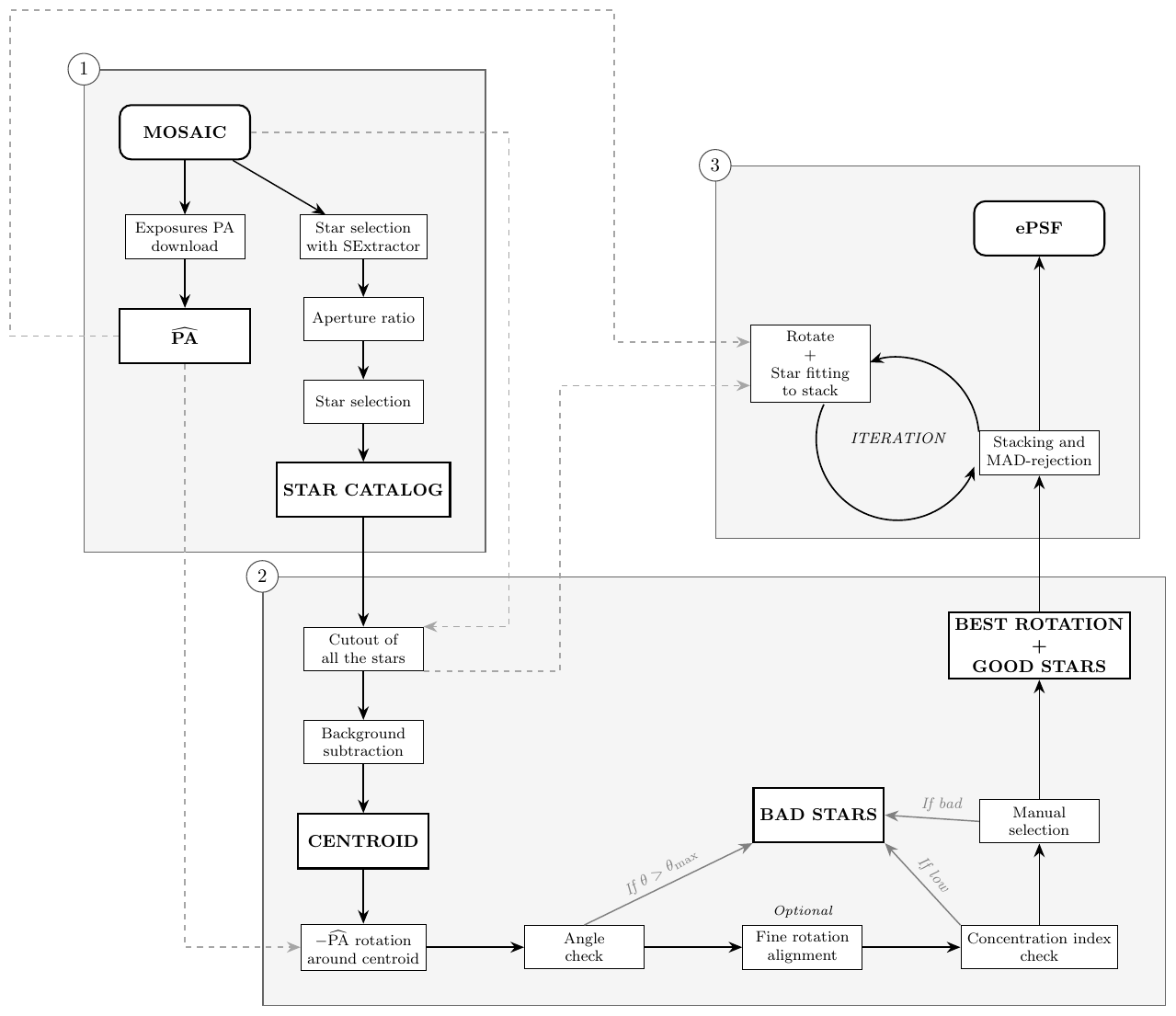}
    \caption{\epi\ pipeline flowchart. \epi\ is composed of three main subprocesses: (1) the creation of the star catalogue and the most common orientation of the input image exposures; (2) the star cutout extraction and the star selection; (3) the stacking.}
    \label{fig:pipeline}
\end{figure*}

We present the \epi\ software, a flexible Python library to easily obtain an average high S/N ePSF from one or multiple science images at once. The software is developed for use with the \textit{James Webb Space Telescope} (JWST) and the \textit{Hubble Space Telescope} (HST) PSFs, though it is generalizable to other telescope images.  \figref{fig:pipeline} illustrates the schematic architecture of \epi, which comprises three main modules. The first one is responsible for star selection and the retrieval of exposure orientations, as discussed in Sects. \ref{sec:pa} and \ref{sec:stars}. The second module performs star analysis to identify candidates for the final ePSF construction, as detailed in \secref{sec:cutouts}. Finally, the third module focuses on the construction of the ePSF, as described in \secref{sec:builder}.

\subsection{Reference orientation for ePSF extraction}\label{sec:pa}
Even though mosaics are typically constructed from exposures taken at a variety of telescope PAs; a single orientation often dominates across the dataset. \epi\ selects stars whose PSF shapes align with the most common exposure orientations, ensuring a consistent ePSF extraction across the field. If the user is not interested in analysing the rotation of the stars, or wants to specify the most common orientation manually, they can simply not invoke this module.

The first step in determining the dominant PA ($\pa$) of a mosaic, which is field and filter-dependent, is to retrieve a list of all observations taken with that filter in the region covered by the mosaic. Assuming the input mosaic is constructed from all available observations, this list corresponds to the individual exposures that make up the mosaic. To automatically determine the orientation of these exposures, \epi\ retrieves metadata via the grizli-cutout web service \citep{grizli:cutout}. The software filters specifically for exposures hosted on the Mikulski Archive for Space Telescopes (MAST) that are published in a fully calibrated format (\texttt{STATUS = 2}). 

\begin{figure}
    \centering
    \includegraphics[width=1\linewidth]{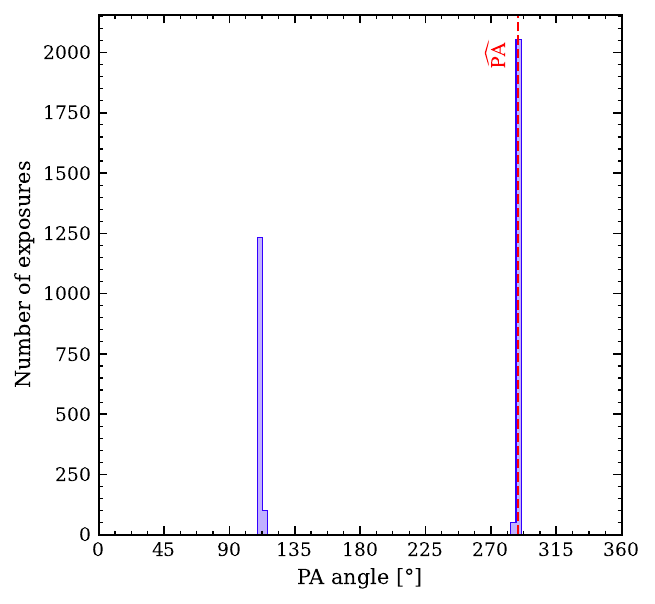}
    \caption{PA histogram of the COSMOS field for the F200W NIRCam exposures. In red, the $\pa$.}
    \label{fig:PA}
\end{figure}

Crucially, the relevant orientation information is stored within the association tables of the targeted exposures. As a result, the software does not need to download the individual, data-heavy exposure files to read their FITS headers, significantly minimising bandwidth and data transfer overhead. From the list of exposure PAs, \epi\ identifies the one with the largest number of exposures within a small tolerance. This most common orientation is what it considers as the mosaic's $\pa$. An example of the PAs of the COSMOS field exposures, obtained with JWST/NIRCam in the F200W filter, is shown in \figref{fig:PA}, along with the selected $\pa$.

A specific case arises when a mosaic contains significant overlaps between exposures taken at different PAs. In such regions, a star may be observed under multiple orientations, resulting in a PSF that is a superposition of these distinct PAs. While \epi\ automatically identifies the primary orientation of the mosaic $\pa$, it does not exclude stars with multiple orientations by default. This design choice is intentional: if a significant portion of the mosaic exhibits multi-PA overlaps --- including the limiting case in which every exposure in the dataset is taken at a different orientation --- these stars can still be included to construct a composite ePSF that more accurately represents the field's PSF, with the rotation-tracking step of the pipeline simply playing no role in the process. Conversely, the user retains full control; through the \texttt{compute\_rotation} function (\secref{sec:cutouts}), one can manually filter the sample to exclude or include stars with different orientations and derive an ePSF tailored to specific scientific requirements.

For science images originating from telescopes or datasets not hosted on MAST, this automated metadata retrieval is unavailable. In such cases, the dominant $\pa$ of the mosaic must be determined independently by the user and provided manually to the \texttt{Image} class.

\subsection{Star selection}\label{sec:stars}
The software builds the ePSF starting from single stars. Therefore, it is crucial to find the position of the unsaturated stars in the field. With this aim, \epi\ makes an external call to the \textsc{SExtractor} software \citep{sextractor:1996, sextractor}, via the \textsc{sewpy} \citep{sewpy:code} package, which retrieves the right ascension and declination of each object, together with their flux and magnitude measured within two apertures of different diameters. By default, the software uses the centroid position of the objects. However, the call to \textsc{SExtractor} is customizable, and the positions determined in this phase do not need to be accurate to subpixel precision, as they will be refined in the subsequent phases.

\begin{figure}
    \centering
    \includegraphics[width=\linewidth]{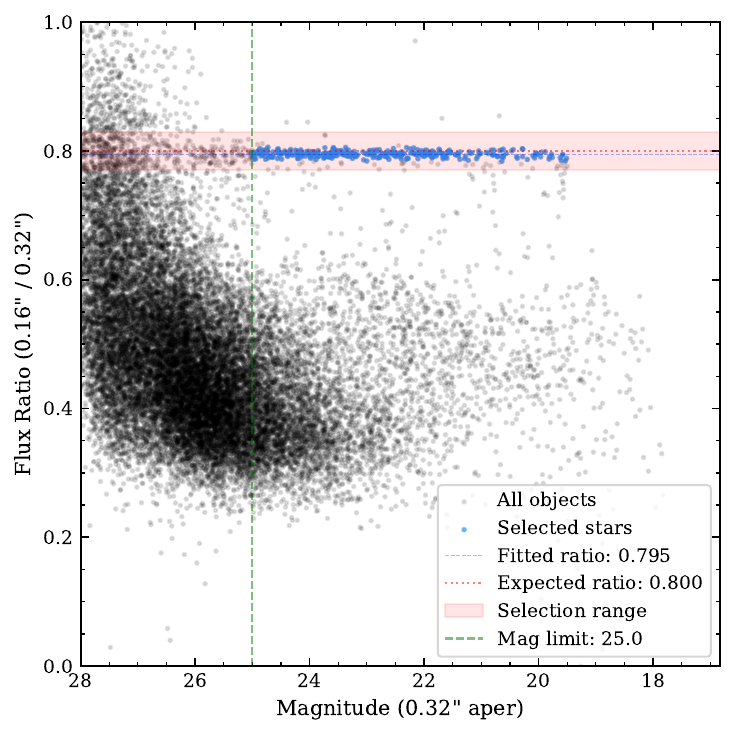}
    \caption{Star selection example in the flux-ratio vs. magnitude diagram. This diagram has been fed with the DAWN JWST Archive (DJA) mosaic of the COSMOS field taken with the F200W filter on the JWST telescope. In black, every object detected by \textsc{SExtractor}; in pink, the manual box set for star detection; in blue, the selected stars that lie on a line at a constant flux ratio. The expected ratio selected by the user is represented by a red horizontal line, while the median flux ratio of the selected stars is visible as a grey horizontal line. The \texttt{mag\_limit}, visible as a green vertical line, is set to 25. This plot shows the exact default output produced by \textsc{epimetheus}.}
    \label{fig:star-selection}
\end{figure}

Following the guidelines provided by \citet{Weaver:2024}, \epi\ adopts default aperture diameters of $0.16\arcsec$ and $0.32\arcsec$. These values are customizable via the \texttt{apertures} parameter within the \texttt{StarSelector} class. 
These aperture sizes are critical for distinguishing point sources from extended objects. Specifically, stars are identified through the flux ratio measured between the smaller (0.16\arcsec) and larger (0.32\arcsec) apertures. Bright, unsaturated stars are isolated in the flux ratio versus magnitude graph, as illustrated in \figref{fig:star-selection} \citep[see also][]{Skelton:2014, Whitaker:2019}.

To select stars with the \texttt{select\_stars} function, the user provides an initial estimate of the flux ratio \texttt{expected\_ratio}) along with a tolerance  \texttt{ratio\_tolerance}). \epi\ computes the flux ratio for all objects detected in the image. Among the objects brighter than a user-specified \texttt{mag\_limit} and whose flux ratio falls within \texttt{expected\_ratio}~$\pm$~\texttt{ratio\_tolerance}, a $k\sigma$-clipping procedure is applied to reject outliers: the median and standard deviation of the flux ratio are computed for the selected sample, and any star deviating by more than $k\sigma$ from the median is discarded. This process is repeated iteratively until convergence, where $k$ is a user-defined parameter.

Figure~\ref{fig:star-selection} shows an example of this selection process, illustrating the initial selection bounds, the fitted flux ratio, and the final sample of selected stars. This diagnostic plot is provided to the user so that the selection criteria can be visually verified.

When extracting a single ePSF from stars across multiple images,  \textsc{SExtractor} is run separately on each one. \epi\ then creates a unique table containing the properties of each selected star.
and the field from which the star was extracted. When multiple stars are located closer than \texttt{r\_star\_separation} (The default value is 0.1 arcseconds, but it can be customized), only one star is retained in the final catalogue: this procedure ensures that stars appearing in overlapping mosaics from the same exposures are not double-counted. Using the flux ratios derived for all sources in the final merged catalogue, \epi\ calculates the mean flux ratio value, $\overline{f_{r}^{\mathrm{SEx}}}$, and the corresponding standard deviation, $\sigma^{\mathrm{SEx}}$.

\subsection{Star selection and analysis}\label{sec:cutouts}
The second module of the software refines the initial catalogue and prepares the data for the ePSF construction. Starting from the previously identified coordinates and magnitudes, this module selects the final sample of stars through an iterative process of automated filtering and user-guided refinement. This module also performs the necessary rotation and alignment for each source. 

\subsubsection{Working of star cutouts}\label{sec:cutout}
\epi\ generates cutouts for each star centred on their respective positions in the image. For each cutout, the local background is estimated using the median flux value after an iterative $k\sigma$-rejection algorithm intended to exclude the star itself and any other nearby object, whenever present.
By default, the cutout size is $\SI{205}{px} \times \SI{205}{px}$. A sufficiently large cutout size is crucial to ensure that the background estimation is unaffected by contamination from the PSF light itself.

A rough estimate of the centroid of each star is obtained by fitting a 2D Gaussian profile to the star, using the \texttt{centroid\_2dg} function from \textsc{photutils}. Each cutout is then mapped to a common grid using the \texttt{drizzle} \citep{Anand:2025} package, centring each star with subpixel accuracy and rotating it by $-\pa$ in a single operation. After this transformation, the star is centred in the cutout. If the star was observed in an exposure taken at $\pa$, with this rotation, it recovers its natural orientation on the sensor, i.e. the orientation of the PSF on the raw exposures before the mosaicking. If \epi\ is used on unmosaicized images, i.e. on the sensor CCD frame, rotation is not computed, and the stars are only shifted to the centre of the cutout.

The drizzling algorithm exploits the sub-pixel centroid positions of individual stars to combine them onto a finer output grid, provided the user requests this via the \texttt{upsample\_factor} parameter. Setting this factor to a value greater than unity yields an ePSF sampled at correspondingly higher resolution than the input images.

\begin{figure}
    \centering
    \includegraphics[width=\linewidth]{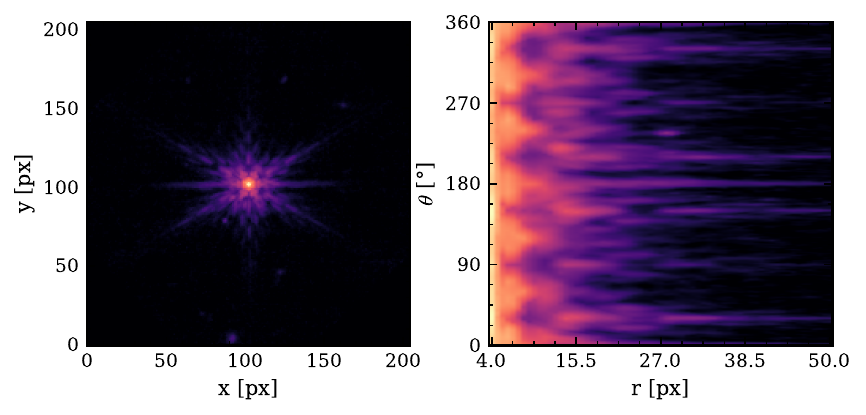}
    \includegraphics[width=0.7\linewidth]{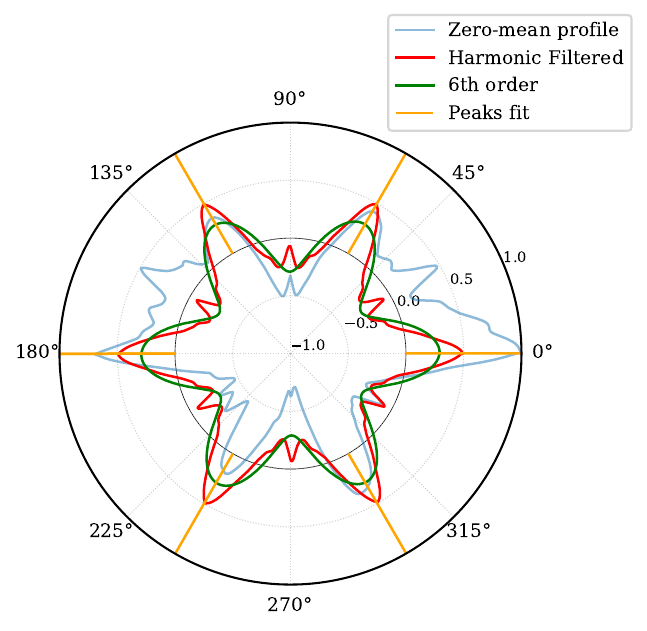}
    \caption{Process to measure the rotation angle of the PSF. \emph{Top} panels: On the left, a single star cutout obtained through the NIRCam F200W filter; on the right, its polar view from \texttt{r\_min=4 px} to \texttt{r\_max=50 px}. \emph{Bottom} panel: the intensity profile of the star PSF after having integrated the polar image in the radial coordinate. In light blue, the normalised zero-mean raw profile; in red, the harmonic-filtered profile, with only harmonics multiple of the sixth; in green, the sixth harmonic; in orange, the fitted position of the peaks.}
    \label{fig:angle-measure}
\end{figure}

If the user prefers to retain only sources aligned with the input image $\pa$ orientation --- a strategy often advantageous in mosaics to obtain a highly representative ePSF on both the center and the wings, as discussed in \secref{sec:discussion} and App.~\ref{app:tests} --- \epi\ provides the \texttt{compute\_rotation} module to automatically quantify the rotation of each star. To achieve this, the software converts each centred and rotated star cutout into a polar representation, spanning radii from \texttt{r\_min} to \texttt{r\_max}, with the radial distance defined from the cutout's central pixel. \figref{fig:angle-measure} illustrates this process for a single source, showing the transformation from Cartesian to polar coordinates. The selection of \texttt{r\_min} and \texttt{r\_max} is critical, as it determines which PSF features contribute to the rotation measurement; these choices are examined in detail in \secref{sec:discussion}, and in general should be tuned on a bright source before applying them to all the stars.

\begin{figure*}[htb]
    \centering
    \includegraphics[width=0.85\linewidth]{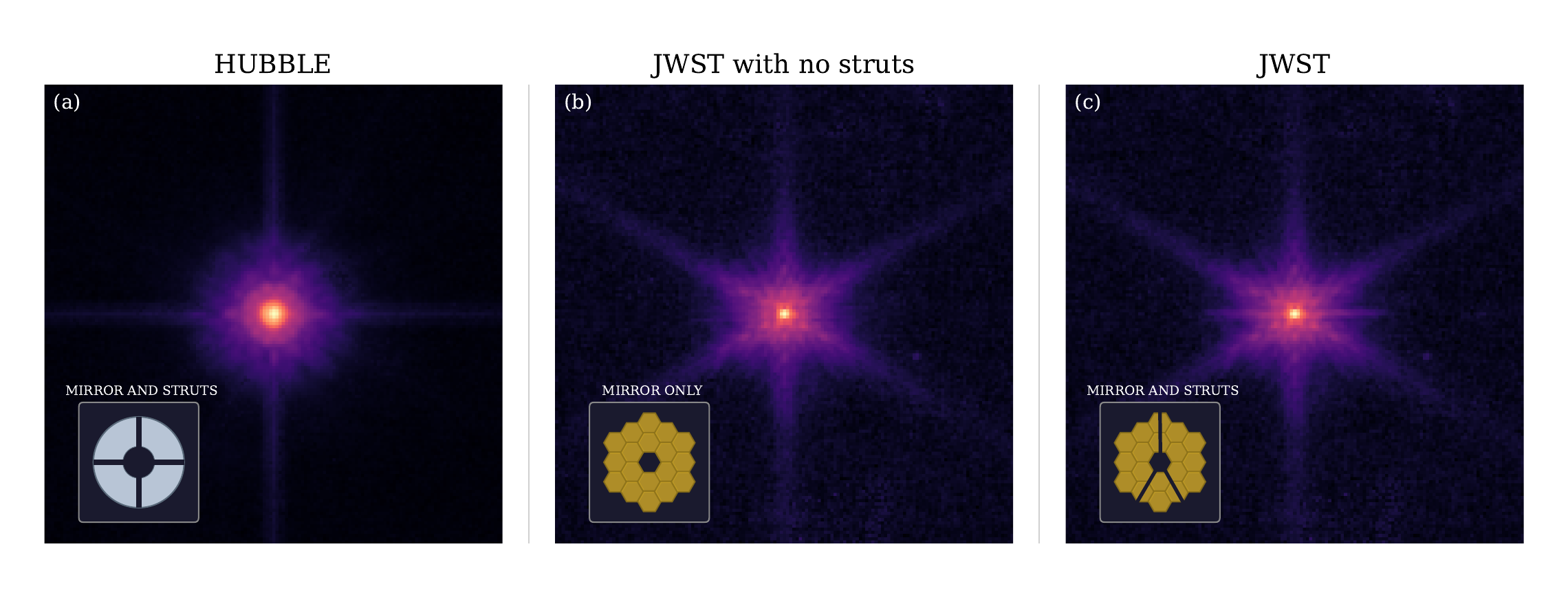}
    \caption{HST F606W PSF and JWST F090W diffraction figures. Each panel show the PSF and, on the bottom left, the shape of the mirror and struts that generate those diffraction figures. Panel (a) shows the F606W ePSF: four spikes are visible at $90^\circ$ one to the other, caused by the telescope struts that hold the secondary mirror in position. Panel (b) is an artistic representation obtained by manually editing the JWST F090W ePSF, which represents the ePSF that JWST would have only because of its hexagonal shape and of each of its mirror tiles. Six spikes separated by $60^\circ$ are produced in this case. Panel (c) shows the complete F090W JWST ePSF caused by both the mirror shape and the structures that support the secondary mirror. Three new spikes are generated by the struts, one of which is horizontal and two that overlap with the spikes generated by the mirror shape.}
    \label{fig:spikes}
\end{figure*}

The polar image is then integrated over the radial direction to obtain an intensity profile of the star's PSF as a function of the angle, and \epi\ performs a fast Fourier transform analysis of the signal. The signal is filtered to retain only the harmonics that are multiples of a specific primary harmonic, determined by the parameter \texttt{nth\_order\_fft}. Indeed, the diffraction figures produced by the mirror shape and the supporting structures of the majority of the telescopes (and in particular on both HST and JWST) produce a periodic signal in the angular direction. For example, HST has cross-shaped spiders that hold the secondary mirror and produce diffraction spikes $90^\circ$ apart \citep{Harvey:1995}, while the JWST hexagonal mirror shape produces 6-point-shaped stars, with the supporting structure of the secondary mirror producing an extra horizontal fainter spike, as shown in \figref{fig:spikes}. Therefore, it is convenient to select \texttt{nth\_order\_fft=4} and \texttt{nth\_order\_fft=6} for HST and JWST, respectively. For other telescopes with non-rotationally-symmetric PSFs, the appropriate value of \texttt{nth\_order\_fft} should be determined independently. The rotation of the PSF is then inferred from the peak positions of the harmonic-filtered signal. 

The user has full control over which star orientations to retain. By default, in line with the original \epi\ concept, the library selects only stars aligned along the most common orientation. This means that after applying a rotation of $-\pa$, these stars remain unrotated with respect to their cutout. The alignment tolerance can be adjusted using the \texttt{angle\_tolerance} parameter within the \texttt{mask\_star\_bad\_rotation} module, specified in degrees. A tighter tolerance results in sharper ePSFs, particularly in the wings, but must be calibrated keeping in mind that a stricter cut reduces the number of available stars, which can compromise the ePSF calibration in star-poor fields. The default value is $2^\circ$. Values below $1^\circ$ are not recommended, as they become comparable to the rotation angle uncertainty. To keep all stars regardless of their rotation, the user can simply not invoke this module. 

\subsubsection{Flux ratio and visual checks}
To ensure flux conservation through the rototranslation step, we verify that the aperture flux ratio (small to large; $f_r$) measured on each transformed stellar cutout, with \textsc{photutils}, remains consistent with the original average \textsc{SExtractor} measurement on stars from the input image $\overline{f_{r}^{\mathrm SEx}}$. Stars with ${f_r}<\overline{f_{r}^{\mathrm SEx}}-k\cdot\sigma^\mathrm{SEx}$ are automatically discarded (with $k = 3$ by default). Low values of $f_r$ at this point of the analysis are likely due to wrongly centred stars or artefacts previously detected by \textsc{SExtractor} as stars.

Stars passing the flux ratio tests are then shown to the user from the brightest to the faintest one by one. The manual selection is fundamental to check the quality of the available cutouts. Interactively, the user can discard any star with an anomalous shape or located in crowded fields, artefacts, and fields with an inhomogeneous background luminosity. When the user is satisfied with the number of selected stars, they can stop the visual check and mark as bad all remaining fainter stars. The minimum number of stars that must be kept strongly depends on the field crowding and the S/N of every single star. From our test, a few tens are generally enough to obtain a high-quality ePSF, as the ones released together with this work. The manual selection and the forced stop of the procedure are managed from a graphical interactive window, working on both Jupyter notebook and pure-Python scripts.

The result of the various selections is always saved to a file, so that running \epi\  again on the same images does not force re-computation of the rotation angle and of the goodness of each star. At this point, a list of good stars, their rough centroids, and cutouts can be passed to the final step of \epi\ to build the ePSF.

\subsection{Building the ePSF}\label{sec:builder}
Creating an ePSF could, in principle, be as simple as performing a noise-weighted stack of the selected stars. However, such a procedure is typically sub-optimal for two primary reasons:
\begin{enumerate}
    \item Centroid alignment alone is insufficient to perfectly align all stellar profiles, leading to an ePSF that is artificially broadened.
    \item Star cutouts frequently contain contaminating sources. For this reason, a robust mechanism to mask unwanted background objects is needed.
\end{enumerate}
To address these challenges, \epi\ employs an iterative, global refinement procedure that simultaneously optimises stellar alignment and pixel-based outlier rejection. The precision of this alignment and the maximum number of iterations are controlled by the user via the \texttt{n\_iterations} and \texttt{convergence\_threshold} parameters in the \texttt{refine\_shifts} function, respectively; typically, five iterations are sufficient to reach stability.

The refinement proceeds as follows: first, all selected star cutouts are normalised by the flux within a default aperture (defined by \texttt{norm\_radius}, set to $0.5\arcsec$ by default). The stellar flux obtained with the aperture photometry is only used to obtain the first iteration of the ePSF; hence, it is not intended to be very precise, and some contamination from other sources can be tolerated. An initial ePSF estimate is then computed by stacking these cutouts on a common pixel grid by using the \textsc{drizzle} package. The drizzling stacking is weighted by the inverse of the variance image, if available, or by the inverse background variance estimated in \secref{sec:cutout}.

This initial stack serves as the ePSF model at the zeroth iteration. In each subsequent iteration, the current ePSF model is rotated to match the orientation of each individual star, accounting for the input image $\pa$; the result is a rotationally aligned ePSF model with respect to the raw stellar cutout. A fitting procedure then determines the optimal shift and flux for each star by minimising the weighted difference between its cutout and the rotated ePSF model. The fitted amplitude is exploited in the next stacking procedure to normalise each star, instead of the aperture photometry used in the initial iteration.

At each iteration, a rejection algorithm is applied to the stacked pixels to mitigate the impact of contaminating sources. The default method is the median absolute deviation (MAD; \citealt{MAD}), which effectively masks outliers and is a particularly robust approach for faint, diffuse objects and when the star sample is small. By varying the \texttt{rejection\_mode} parameter, the user can select between \texttt{"mad"} (the default), \texttt{"weighted"} (a noise-weighted $k\sigma$ rejection), or \texttt{"none"} (disabling rejection). Following outlier rejection, the unmasked pixels are weighted by their inverse variance and stacked to produce a refined ePSF estimate. This process is repeated until the root mean square of the corrections in pixels is lower than \texttt{convergence\_threshold} or until the specified \texttt{n\_iterations} number of iterations are reached, at which point the final centroids and the resulting ePSF are adopted.

Once these centroids are established, \epi\ can create the final ePSF for each image from which we extracted the stars, each potentially having a different original $\pa$. Indeed, the software can derive an ePSF optimised for each image separately, even when multiple images are loaded simultaneously to increase the number of available stars. To achieve this, \epi\ repeats the complete stacking procedure --- centring, rotation, outlier rejection, and stacking --- after rotating each star cutout to align with the target $\pa$ of the specific image. During this phase, the user can choose to force a perfect alignment of all stellar profiles via the \texttt{fine\_rotation = True} parameter. When enabled, \epi\ rotates the cutouts by accounting not only for the macroscopic target $\pa$, but also by adding the individual rotation angle of each star, as computed by the \texttt{compute\_rotation} module described in \secref{sec:cutout}. This behaviour is disabled by default, as it may introduce some biases if the rotation is not computed correctly, as further discussed in  \secref{sec:discussion}.

This stacking process can be performed for any desired target $\pa$, allowing multiple ePSFs to be generated from the same set of stars, each corresponding to a different orientation. Importantly, each ePSF is built directly from the original star cutouts, meaning every star undergoes only a single rototranslation. This approach minimises the interpolation and blurring effects that would otherwise arise from repeated geometric transformations. Furthermore, because the stars preserve the information of their positions inside the original images, \epi\ can be used to obtain different rotated ePSFs for distinct regions of the field. This allows the model to better align with the specific local rotations of the exposures at different image positions.

\subsection{ePSF quality statistics}
To assess the quality of the retrieved ePSF, \epi\ obtains two parameters: quality-of-fit ($q_f$) and sharpness ($C$). Both parameters are defined according to the ones used in the hst1pass software \citep{Anderson:2022}. In particular, $q_f$ is the sum of the absolute value of the residual of the PSF-fit over a standard $5\times5$ aperture divided by the measured flux, while $C$ measures the fractional flux excess in a star’s central pixel:
\begin{align}
    q_f&=\frac{\sum_{5\times5}\bigl|{I-B-a\cdot \psi}\bigr|}{a}\\
    C&=\frac{{(I-B-a\cdot \psi)_{\rm centroid}}}{a}
\end{align}
with $I$ representing the star cutout, $B$ denoting the background, $\psi$ being the shifted ePSF fitted to the star, and $a$ signifying the fitted star amplitude. The value of $q_f$ is calculated by summing the absolute values of the residuals within a $5\times5$ pixel region centred on the star's centroid, and this sum is normalised by the star's fitted flux. In contrast, sharpness is computed solely for the centroid pixel and does not use absolute values.

\begin{figure}
    \centering
    \includegraphics[width=\linewidth]{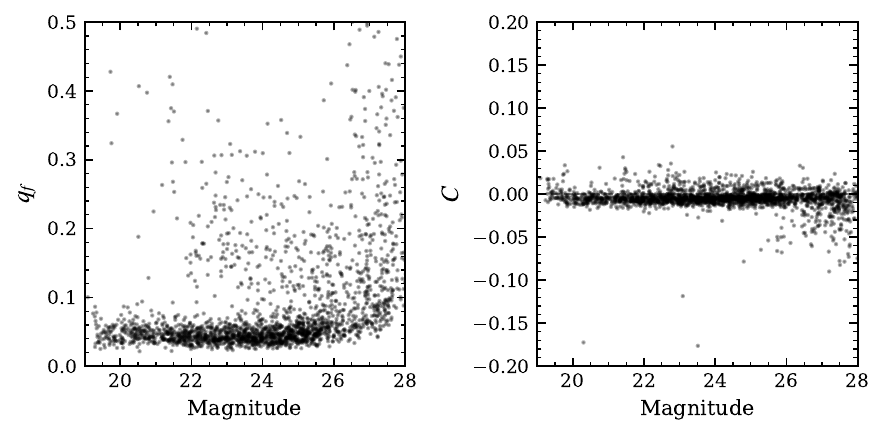}
    \caption{Statistical adimensional parameters $q_f$ and $C$ for all the objects up to magnitude 28 for the F200W NIRCam exposures of the Abell 2744, COSMOS, UDS, and GOODS-S fields together. This plot shows the exact default output produced by \textsc{epimetheus}.}
    \label{fig:statistics}
\end{figure}

The size of the $q_f$ region can be customised using the \texttt{qFit\_region\_size} parameter. The values of $q_f$ and $C$ for all selected (\secref{sec:stars}) stars are presented in \figref{fig:statistics}. Generally, objects with a $q_f$ value higher than $\gtrsim0.2$ are likely artefacts, resolved objects, stars that are prospectively close to one another, or very low S/N stars, indicating in general poor fitting. Additionally, at lower magnitudes, the $q_f$ value tends to naturally increase due to the decreased S/N.

\begin{figure}[htb]
    \centering
    \makebox[0.28\linewidth][l]{\tiny\textbf{Star cutout}}%
    \makebox[0.28\linewidth][c]{\tiny\textbf{Fitted ePSF}}%
    \makebox[0.28\linewidth][r]{\tiny\textbf{Residuals}}\par
    \foreach \i in {0,...,5}{%
        \includegraphics[width=\linewidth]{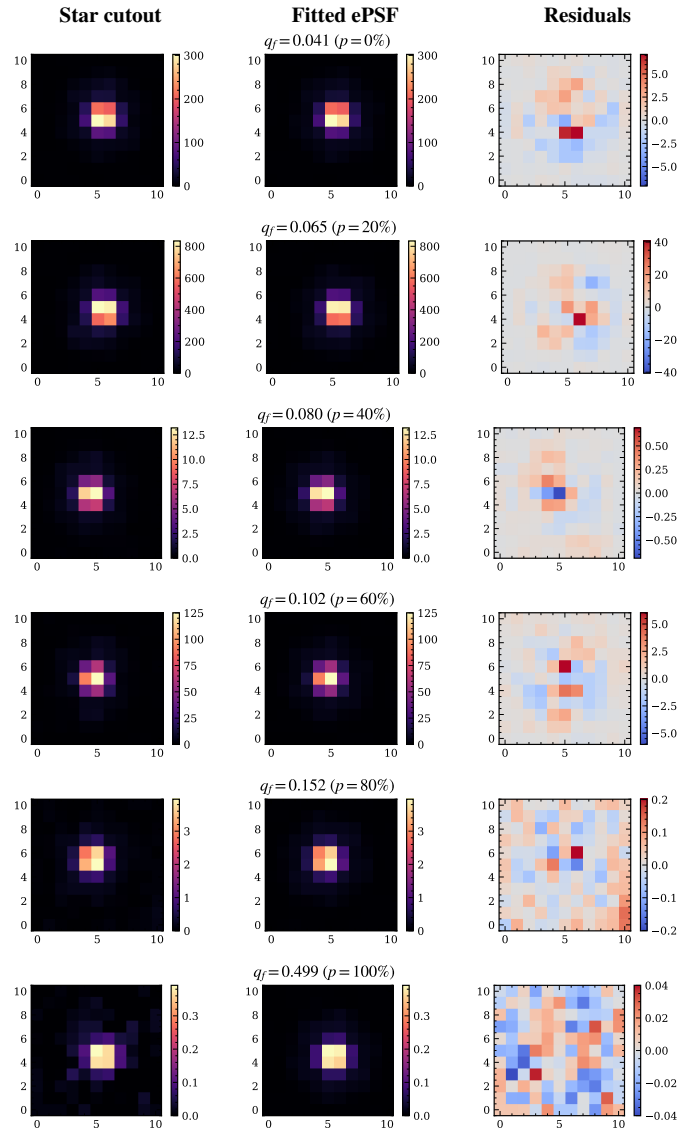}
    }
    \caption{Star cutouts (\emph{left}), fitted ePSF to the star (\emph{centre}), and residuals (\emph{right}) for stars with different quality fit values. In each graph, $p$ indicates the percentile of the $q_f$ of the star represented, among all the stars in the F200W filter ot the the Abell~2744, COSMOS, UDS, and GOODS-S fields with a magnitude lower than 28.}
    \label{fig:residuals}
\end{figure}

To further assess the quality of the fits, \figref{fig:residuals} shows the residuals between the star cutouts and the corresponding fitted ePSF for a selection of six stars spanning different values of $q_f$, with cutouts of $11\times11$ pixels; the stars are chosen at the $0\%$, $20\%$, $40\%$, $60\%$, $80\%$, and $100\%$ percentiles of the $q_f < 0.5$ distribution shown in \figref{fig:statistics}, computed from all stars across the Abell~2744, COSMOS, UDS, and GOODS-S fields combined, and in general no clear patterns are visible in the residuals (also when inspecting a larger sample of stars), indicating that the fits are good and that the residuals are dominated by noise.

\subsubsection{Quality tests and ePSF release}\label{sec:data}
We use the data from four well-studied extragalactic fields to test the \epi\ performances: Abell 2744, COSMOS, UDS, and GOODS-S. These fields benefit from extensive multi-wavelength coverage from both the HST and the JWST, spanning a wide range of filters from the optical to the near-infrared. All mosaics used in this work are retrieved from the DAWN JWST Archive and reduced using the \textsc{grizli} software \citep{https://doi.org/10.5281/zenodo.1146904}, which provides uniformly reduced imaging products drizzled to a common 20~mas or 40~mas pixel scale, depending on the filter. A full list of programs contributing to these mosaics is provided in the acknowledgements of this work.

\begin{figure*}
    \centering
    \includegraphics[width=0.75\linewidth]{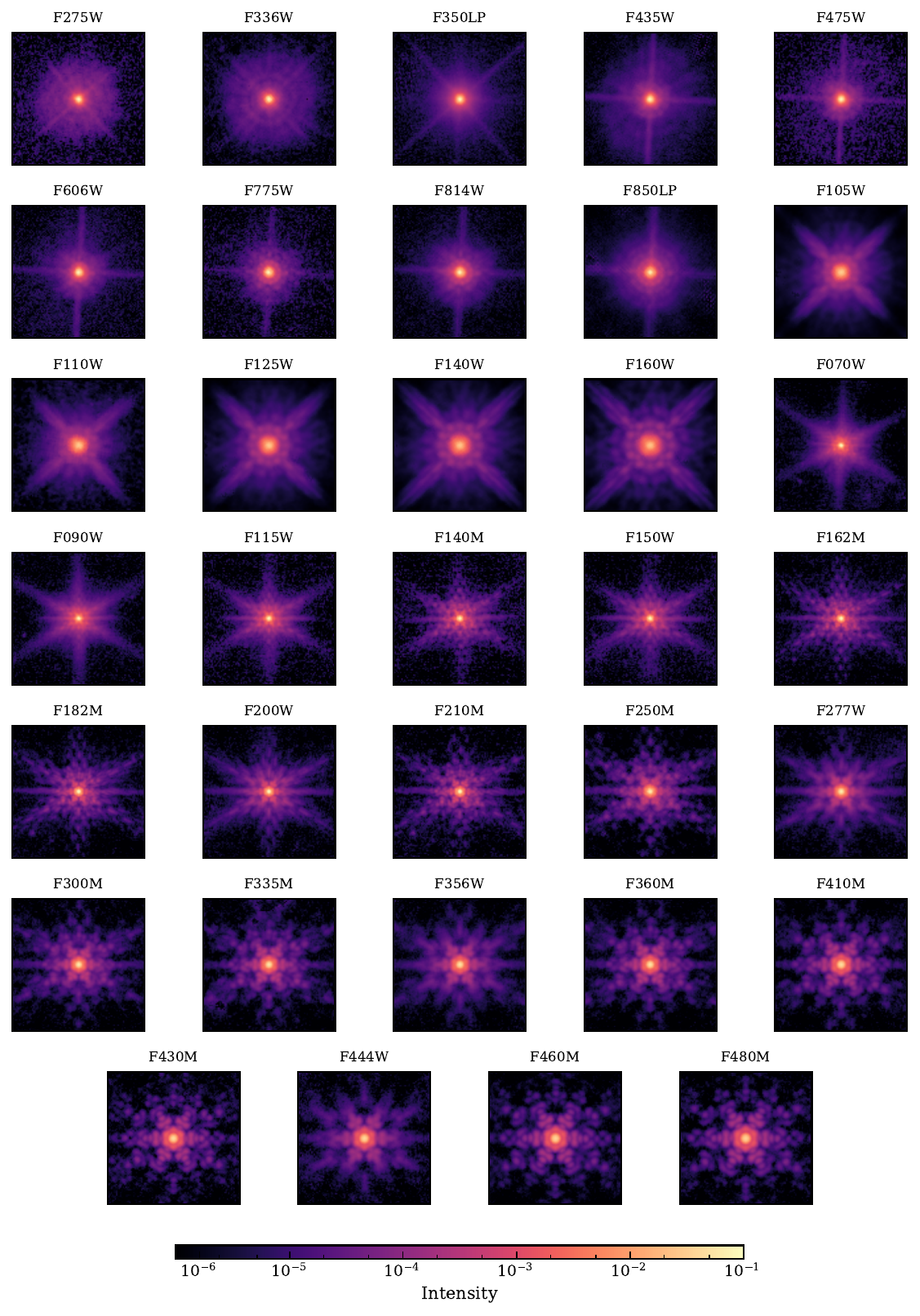}
    \caption{ePSFs retrieved with \epi\ on the Abell 2744, UDS, COSMOS, GOODS-S fields simultaneously. The displayed ePSFs are oriented as they are naturally oriented on the sensor.}
    \label{fig:all-psfs}
\end{figure*}

We computed the ePSF for all available HST and JWST NIRCam filters using the data from the COSMOS, UDS, GOODS-S and Abell 2744 fields simultaneously. These ePSFs are visible in \figref{fig:all-psfs} and can be downloaded from the \epi\ webpage, together with the \epi\ package. The growth curves of the computed ePSFs are discussed in Appendix~\ref{app:growth}, both before any convolution and after the convolution of each ePSF to the F480M one, to simulate the performance of these ePSFs in photometric measures.

Detailed quality tests and statistical diagnostics demonstrating the accuracy and reliability of \epi\ are provided in Appendix~\ref{app:tests}. These results offer valuable insight into the software's performance across a variety of observational conditions and input images configurations.

\section{Discussion}\label{sec:discussion}
\epi\ is a comprehensive software designed to create an ePSF from images, including the option to work with multiple input images simultaneously. The software handles all necessary components, such as object identification, star selection, star rotation angle measurement, subpixel alignment, rejection of bad pixels, and stacking, and is easily customizable for the user's needs.

To obtain the best possible result, a variance image is required, together with the science one. This additional image can be used by both \textsc{SExtractor} to identify objects and by \epi\ itself to perfectly weight the stacking procedure. The variance image, which can also be the root mean square image or a weight image, must have the same world coordinate system as the science image. If such an additional image is not provided, the stacking procedure is computed by evaluating the standard deviation of the background of each cutout, but the procedure is less accurate as it cannot take into account the spatial variation of the noise inside each cutout.

\epi, in its current version, is designed to compute the average ePSF of the image exactly as it is provided to the software. This means, in particular, that the position retrieved for each star is simply its position in the image coordinate system, and the sky coordinates of the stars are those retrieved from \textsc{SExtractor}. No accurate distortion model of the image is considered, but the image can be passed to \epi\ after being corrected for geometric distortion. For this reason, the retrieved ePSF is accurate for the input image as provided, and the star positions are extremely accurate only in the image frame.

\epi\ also offers the option to produce an upsampled image during the processing. This upsampling is performed alongside the rototranslation to ensure that only a single transformation is applied to each cutout. Once the final upsampled ePSF is computed, \epi\ applies binning to retrieve the ePSF at standard resolution. Both the upsampled and the standard resolution ePSFs are outputs of the \epi\ pipeline. By default, an upsampling factor of four is used, but can be user-customised.

\epi\ also supports input images with different spatial resolutions, provided the pixel scales are commensurate (i.e., integer multiples of each other, such as 0.02, 0.04 and 0.08\arcsec/px). This situation typically occurs, for example, with the short-wavelength filters of JWST, where mosaics reduced with \textsc{grizli} are released at resolutions of both 0.02\arcsec/px and 0.04\arcsec/px. In such cases, it is possible to combine images from both resolutions to create an ePSF using the lowest resolution among the images. \epi\ automatically detects the different resolutions, eliminating the need for user adjustments.

Spatial and temporal variations of the PSF \citep{Anderson:2000, Nardiello:2022, Zhuang:2024} may limit the accuracy of the ePSF obtained using the \epi\ approach. Temporal variations are, in our view, the most difficult to mitigate in mosaics where exposures taken at very different epochs are often stacked together; the best one can realistically aim for is an effective, time-averaged ePSF that best represents the mosaic as a whole. Regarding spatial variations, the current version of \epi\ does not include a dedicated pipeline to produce a spatially varying ePSF across the field of view; however, each star is represented in the code by an object of the \texttt{StarContainer} class, which retains the star's detector position together with all other measured properties. As a result, it is already possible to restrict the star sample to a chosen sub-region of the mosaic and derive an ePSF for that region alone, repeating the procedure over multiple sub-regions to build up a spatially dependent set of ePSFs. We plan to implement this as an automated, built-in feature in a future release of \epi.

\subsection{Spikes direction and star rotations}
One of the main advantages of \epi\ is the automatic measurement of the rotation angle of the stars. To work properly, this method must first be tuned with prior information about the ePSF. In particular, the user must provide the expected position, though the \texttt{angle\_of\_one\_spike} parameter, of one of the spikes as seen in the raw images on the sensor field. That means that the user must have a minimum knowledge of the instrument used and the sensor orientation in the focal plane. For example, the HST spikes are rotated by $\sim 45^\circ$ in the WFC3 images with respect to the one taken with ACS camera \citep{ACS_Handbook:2025, WFC3_Handbook:2025}, which means that the spike angles are different with different instruments.

For JWST, the structure of the PSF spike is quite complex, and the exact initial angle depends on the combination of factors such as the filter used, the polar \texttt{min\_radius}, and the image resolution. A recommended set of parameters for each filter is provided with \epi; however, the optimal combination may vary from one case to another. Therefore, when working with telescopes different from the HST and JWST, users need to test the parameters on a few stars before applying the full analysis. This ensures that the angle detection is functioning correctly.

The use of the \texttt{fine\_rotation = True} parameter, which forces each star to be aligned with the other, may be useful in extreme cases where multiple PAs are present in the same image. However, this function may introduce some biases in the final ePSF as each star is exactly aligned to \texttt{angle\_of\_one\_spike}. However, in the case of low S/N stars, it might be useful when retrieving ePSF spike structures from mosaics built from exposures collected with a wide range of PA exposures.

\section{Conclusions}\label{sec:conclusions}
We present \epi, a customizable Python-based software designed to extract ePSFs from astronomical images and mosaics. The \epi\ pipeline operates by selecting stars across one or more fields simultaneously, determining the most common orientation of the exposures comprising the input image, fitting the rotation angle of each star's PSF, and either selecting stars with a specific orientation or rotating all stars to a common direction, or both. Finally, \epi\ requires manual inspection of each stellar candidate, avoiding restrictive automated cuts. Such hands-on verification can simultaneously maximise the stellar sample and reliably exclude artefacts and contaminated sources. Through iterative subpixel alignment and stacking, the software produces a high-quality ePSF that accounts for the rotational characteristics of the input data.

\epi\ can be used:
\begin{itemize}
    \item To retrieve a mean ePSF for different mosaics obtained with the same filter using all the available stars together. This pipeline procedure produces very high S/N ePSFs, each one rotated as the $\pa$ of each field, which may lead to better results than using the stars of each field to get the ePSF of that field only
    \citep{Merlin:2024}.
    \item To deal with a mosaic composed of exposures with a variety of PAs. \epi\ allows fitting and rotating each star to the same direction, to obtain a high-quality ePSF even when few stars are exactly co-aligned. If the stars are stacked together without having them co-rotated, the rejection methods could cancel out the outer part of the ePSF while stacking, resulting in artificially higher flux in the central region when the final ePSF is normalised, as discussed in Appendix \ref{app:tests}.
    \item To easily identify the best ePSF for an image, even when angle fitting or rotation are useless, as \epi\ does not impose unnecessary steps. For example, if all the exposures are aligned in the same direction, and so all the stars have co-aligned spikes, the affine transformation used to rototranslate each star will not rotate anything, and it will only shift the cutouts to align the stars at the subpixel level.
\end{itemize}

Alongside the release of the \epi\ library, we provide ePSF models derived for four widely-used extragalactic survey fields: COSMOS, Abell 2744, UDS, and GOODS-S. For each available HST and JWST NIRCam filter, we supply both the unrotated ePSF and a version rotated to match the most common exposure orientation for each field.

\epi\ and the accompanying ePSF library will serve as valuable resources for the astronomical community, facilitating photometric measurements, PSF matching, and image analysis across a wide range of fields and surveys, thanks to its customizable workflow. In the context of the extragalactic field, the ePSF provided by our pipeline would be useful for a variety of applications, such as deriving unbiased structural parameters and Sérsic profiles for faint and extended galaxies, decoupling the nuclear emission of Active Galactic Nuclei (AGN) from their hosts across multiple pointings, and performing accurate photometry of faint background systems (such as ultra-diffuse galaxies or compact star clusters embedded within dense galaxy clusters).

\begin{acknowledgements}
We are grateful to the anonymous referee for their valuable comments and suggestions, which have helped improve the manuscript. We are grateful to E. Merlin for the fruitful discussions that inspired the development of this work, and to P. Watson and A. Milone for their insightful comments and suggestions, which significantly improved and optimised \epi. 

This research is based on observations made with the NASA/ESA Hubble Space Telescope obtained from the Space Telescope Science Institute, which is operated by the Association of Universities for Research in Astronomy, Inc., under NASA contract NAS 5–26555. These observations are associated with programs:
\#11689 (PI: Dupke), \#13386 (PI: Rodney), \#13495 (PI: Lotz; HFF), \#13389 (PI: Siana), \#15117 (PI: Steinhardt; BUFFALO), \#17231 (PI: Treu),
\#9822 and \#10092 (PI: Scoville; COSMOS Treasury), \#12440 and \#12443 (PIs: Faber \& Ferguson; CANDELS),
\#12177 and \#12328 (PI: van Dokkum; 3D-HST),
\#9425 and \#9583 (PI: Giavalisco; GOODS), \#11563 and \#12498 (PI: Illingworth; HUDF), \#12534 (PI: Teplitz; UVUDF).

This work is also based on observations made with the NASA/ESA/CSA James Webb Space Telescope. These observations are associated with programs:
GO-2561 (UNCOVER; PIs: Labb\'e \& Bezanson), GO-4111 (MegaScience; PI: Suess), ERS-1324 (GLASS; PI: Treu), DD-2756 (PI: Chen), DD-2767, GO-2883 (MAGNIF; PI: Sun), GO-3516 (ALT; PIs: Naidu \& Matthee), GO-3538 (PI: Iani),
GO-1837 (PRIMER; PI: Dunlop), GO-1727 (COSMOS-Web; PIs: Kartaltepe \& Casey),
GTO-1180 (JADES; PI: Eisenstein), GTO-1210 (JADES; PI: L\"utzgendorf), GO-1895 (FRESCO; PI: Oesch), GO-1963 (JEMS; PI: Williams).

The raw data were obtained from the Mikulski Archive for Space Telescopes at the Space Telescope Science Institute, which is operated by the Association of Universities for Research in Astronomy, Inc., under NASA contract NAS 5-03127 for JWST. The processed data products were retrieved from the DAWN JWST Archive (DJA, \url{https://dawn-cph.github.io/dja/}). DJA is an initiative of the Cosmic Dawn Center (DAWN), which is funded by the Danish National Research Foundation under grant DNRF140.
\end{acknowledgements}
\bibliography{aa}

\appendix\onecolumn\nolinenumbers
\section{ePSF growth curves}\label{app:growth}
\begin{figure*}[hb]
    \centering
    \includegraphics[width=1\linewidth]{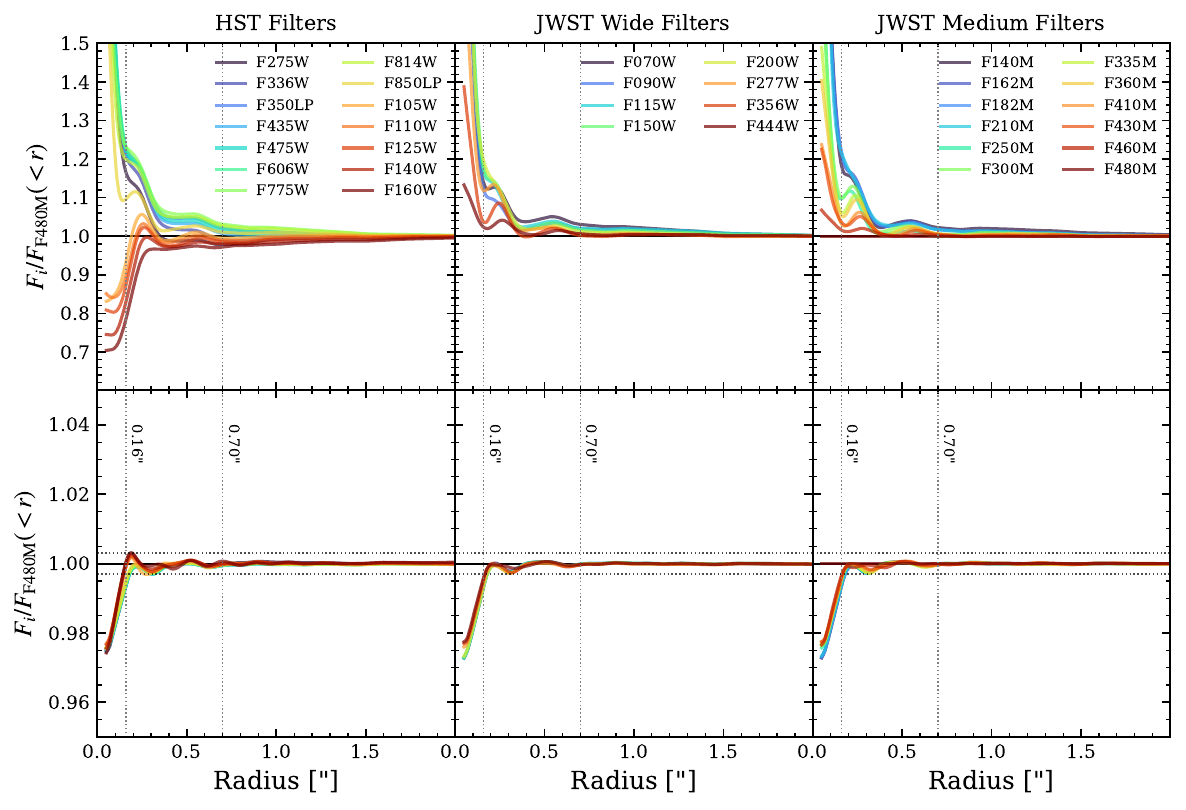}
    \caption{Growth curves of the ePSFs for HST (left panel), JWST NIRCam wide (centre), and JWST NIRCam medium (right) filters, normalised to the F480M. The top panels show the growth curves of the unconvolved ePSFs, while the bottom ones show the growth curves after convolution to match the F480M ePSF. The horizontal dotted lines in the bottom panels indicate $\pm0.3\%$ deviations from unity. Vertical dotted lines mark aperture radii of 0.16\arcsec and 0.70\arcsec.}
    \label{fig:growth}
\end{figure*}

We computed the growth curves of the retrieved ePSFs for HST and JWST NIRCam filters. The growth curves, normalised to the F480M, which is the widest NIRCam filter, are shown in \figref{fig:growth} for the ePSFs oriented as they naturally appear on the detector frame. We present both the growth curves of the unconvolved ePSFs and those convolved to match the F480M filter. The convolution kernels are obtained using \textsc{pypher} \citep{boucaud2016} with a regularisation parameter of $3\times10^{-3}$ on $3\times$ upsampled ePSFs, after which the kernels are reduced to the original pixel scale. This procedure follows that of \citet{Weaver:2024}.

Thanks to the very high S/N of the ePSFs obtained using \epi, we are able to derive convolution kernels that are not dominated by noise for wide, medium, and HST filters, even using low regularisation parameters in \textsc{pypher}. In the convolved case, the maximum deviation among the growth curves reaches $0.5\%$ at just $0.16\arcsec$ from the centre, demonstrating that the PSF homogenization is highly effective across all filters. This level of agreement ensures that when aperture photometry is performed with the \epi\ ePSFs, it is possible to reach sub-percent accuracy, a critical requirement for robust spectral energy distribution fitting and photometric redshift estimation. At larger radii ($>0.7\arcsec$), the convolved growth curves converge to within $\sim0.1\%$ of unity, indicating excellent flux conservation in the outer wings of the PSF.

\section{ePSF quality tests}\label{app:tests}
\begin{figure*}[hbp]
    \centering
    \includegraphics[width=0.75\linewidth]{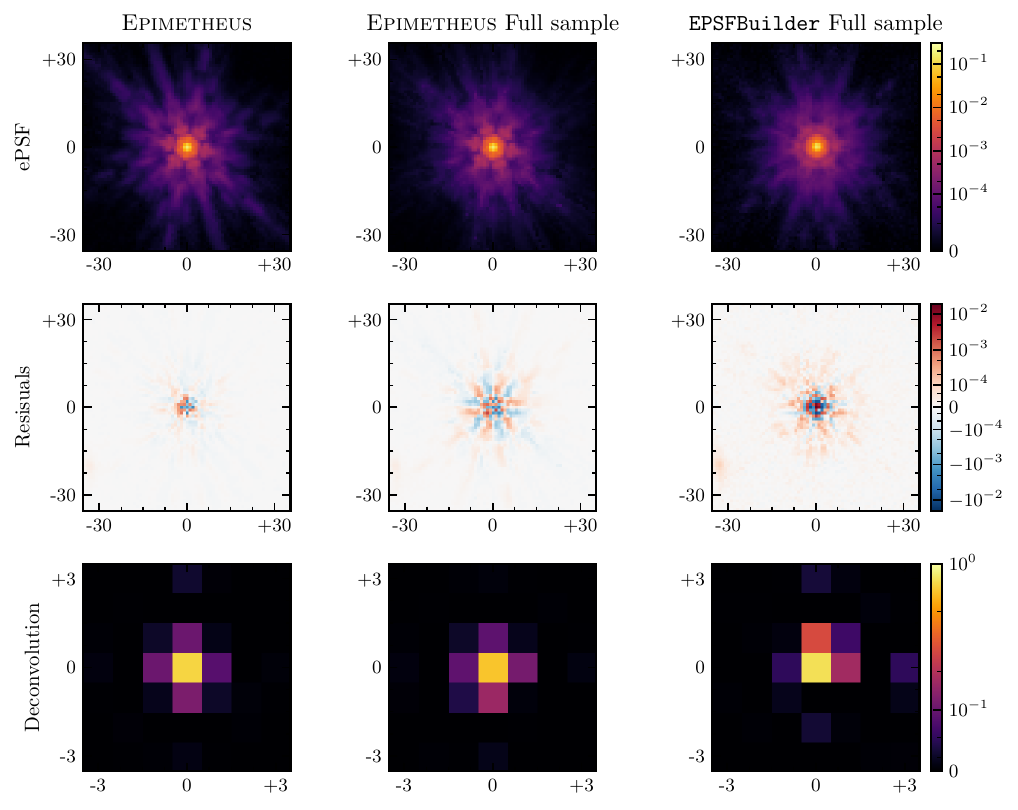}
    \caption{ePSF performances with different software and settings applied on the F200W COSMOS, UDS, Abell 2744 and GOODS-S fields simultaneously. Left column: The ePSF obtained with \epi\, as described in this article; the ePSF shown is the one oriented along the COSMOS $\pa$. Middle column: the ePSF obtained with all the stars regardless of their orientation, so the stars are stacked without checking they are oriented as the field $\pa$; the ePSF shown is the one oriented for the COSMOS field. Right column: the resulting ePSF obtained using \texttt{EPSFBuilder} function from the \textsc{photutils} package. For each column, the first row shows the ePSF, the middle row the stacked residuals between the normalised stars and the aligned ePSF to the star itself, while the bottom row shows the stack of the aligned deconvolved images using 500 RL steps. Both the residuals stack and the deconvolution stack are normalised by the number of stars.}
    \label{fig:performance-multiple}
\end{figure*}

To test the efficiency of \epi, we perform some quality tests on two different scientific cases.
In the first one, the goal was to retrieve a common ePSF for the F200W COSMOS, UDS, Abell 2744 and GOODS-S fields simultaneously. Initially, we extracted the 200 brightest stars among all those extracted in the four mosaics and manually masked those having other bright objects in the field or with irregular background luminosity. This first part of the processing is made using \epi\, and a final sample of {\color{red}} stars with varying orientation is produced. We compute the average ePSF from these cutouts in three ways:
\begin{itemize}
    \item Using \epi\ to obtain four different ePSF, one per mosaic, generated using all the stars from all mosaics, but obtaining the final ePSF oriented as the $\pa$ of each mosaic.
    \item Using \epi\ to obtain a single ePSF using stars from all the fields without ever rotating the stars and stacking them together as they are.
    \item Using the \texttt{EPSFBuilder} class from the \textsc{photutils} package to build a single ePSF using all the stars in the fields. 
\end{itemize}
\figref{fig:performance-multiple} shows the obtained ePSF in the three cases together with the cumulative residual of the retrieved ePSF subtracted from each centred and normalised star. The same figure also shows the stack of the deconvolved stars using the obtained ePSFs. The deconvolution was computed using 500 steps in the Richardson-Lucy algorithm using the \textsc{scikit-image} \citep{scikit-image} package.

The results clearly show how the residuals in the wings are better when using the ePSF oriented as $\pa$ for each mosaic, while the ePSF obtained with $\texttt{EPSFBuilder}$ has too faint wings, as the residuals are consistently positive in a circular corona from 10~px to 20~px far from the centre. The difference between the unrotated \epi\ ePSF and the $\texttt{EPSFBuilder}$ one comes from the fact that the latter automatically chooses a subsample of stars to build the ePSF, and at least in this test, the choice is not optimal. The majority of stars in our sample come from the COSMOS field, and for this reason, the \epi\ unrotated ePSF almost matches the COSMOS $\pa$ rotation. Moreover, the rejection methods implemented by default in the two algorithms are different: \epi\ uses the MAD rejection, while $\texttt{EPSFBuilder}$ uses a $k\sigma$-rejection.

Furthermore, the core of the ePSF is sub-optimally modelled by \texttt{EPSFBuilder}, a limitation that is clearly evidenced by the residual map. Although the deconvolution process recovers a compact central source in all three cases, the results produced using the \epi\ ePSFs exhibit a noticeably higher degree of symmetry. This superior symmetry indicates a more accurate mean reconstruction of the stellar profile, which ultimately translates into reduced systematic uncertainties in subsequent analyses.

This initial test serves a dual purpose: it demonstrates the robustness of \epi\ when processing multiple images simultaneously, and it simulates the realistic scenario of deriving an average ePSF from a single, complex mosaic constructed from exposures with widely varying PAs. The results confirm that \epi\ successfully recovers a high-fidelity ePSF that provides a superior description of the true stellar profiles. Furthermore, in such complex mosaic scenarios, \epi\ could be used to allow the user to make use of the entire stellar sample to generate a suite of orientation-specific ePSFs. These tailored models can subsequently be applied to distinct regions of the mosaic, ensuring that the PSF model accurately matches the dominant orientation of the exposures in any given area of the field.

The second quality test is performed on a subsample of stars of the COSMOS field that have all the same orientation (within a $2^\circ$ tolerance), to mimic a field where every exposure has the same PA.

In this test, three different ePSFs are built:
\begin{itemize}
    \item An ePSF obtained through \epi, using stars coming not only from COSMOS, but also from Abell 2744, UDS and GOODS-S. The final ePSF is obtained by stacking all the stars, orienting them in the direction of the stars used in this test.
    \item An ePSF using only the stars from COSMOS with the same orientation. In this case, no rotation is performed as all the stars already have the same rotation.
    \item An ePSF built using \texttt{EPSFBuilder} on the same stars.
\end{itemize}

\begin{figure*}[ht]
    \centering
    \includegraphics[width=0.75\linewidth]{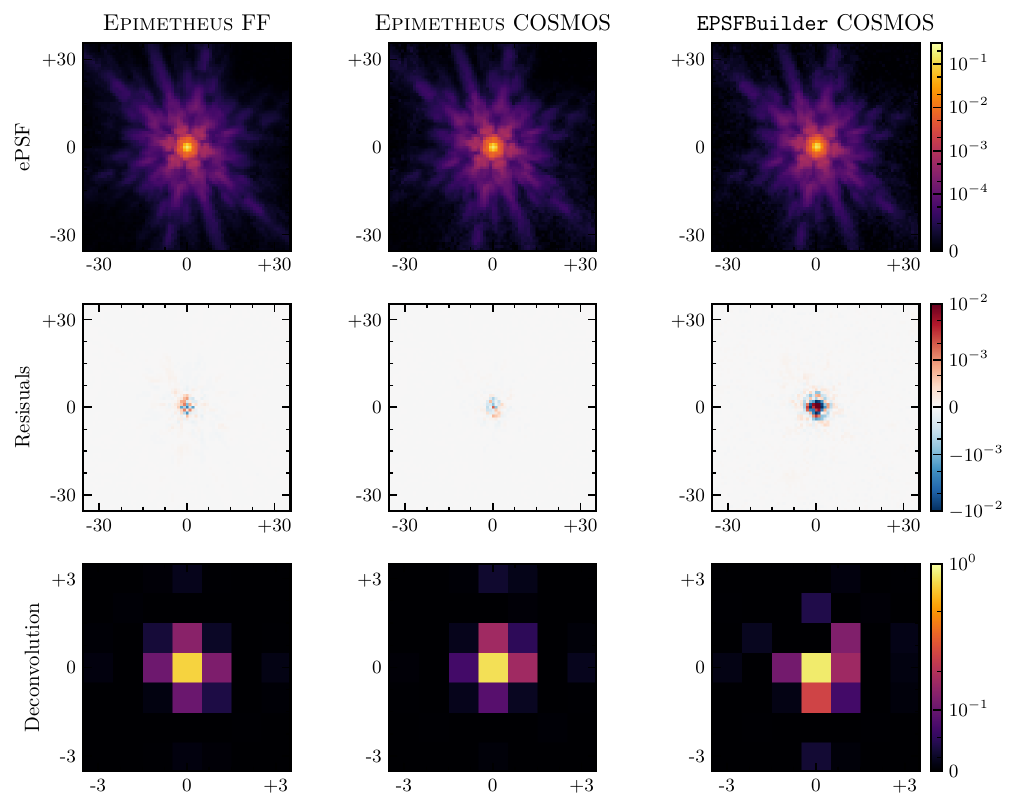}
    \caption{Same as \figref{fig:performance-multiple} for a subsample of stars of the F200W COSMOS field with the same orientation. Left column: The ePSF obtained with \epi\, as described in this article, using the COSMOS, UDS, Abell 2744 and GOODS-S stars. Middle column: the ePSF obtained with \epi\ using the selected COSMOS stars only. Right column: the resulting ePSF obtained using \texttt{EPSFBuilder} function from the \textsc{photutils} package.}
    \label{fig:performance-direction}
\end{figure*}

Figure \ref{fig:performance-direction} displays the obtained ePSFs, the residuals between the stars and the ePSF and the RL deconvolution results. In this comparison, the results from the two \epi\ ePSFs highlight the inherent trade-off between sample size and field specificity. As expected, the ePSF constructed exclusively from the COSMOS field yields the lowest residuals, since the model is evaluated on the exact same stellar sample used for its derivation. 
Conversely, the composite ePSF built simultaneously from multiple mosaics exhibits slightly higher residuals due to the intrinsic morphological variance introduced by combining stars across different observational epochs. However, this multi-mosaic ePSF yields a significantly more azimuthally symmetric RL deconvolution. This improvement is driven by the higher S/N achieved through the larger stellar sample, which effectively mitigates the well-known tendency of the RL algorithm to overfit and amplify noise artefacts at high iteration counts \citep[e.g.,][]{Liu:2025}. Ultimately, this test demonstrates that the best ePSF choice may depends on the specific science goal. \epi\ provides the necessary flexibility to align with the user's needs, depending on the specific requirements of their scientific analysis.

\label{LastPage}
\end{document}